\documentclass{aa}  
\usepackage{graphicx}
\usepackage{txfonts}
\def\simlt{\lower.5ex\hbox{$\; \buildrel < \over \sim \;$}}
\def\simgt{\lower.5ex\hbox{$\; \buildrel > \over \sim \;$}}
\def\simpropto{\lower.2ex\hbox{$\; \buildrel \propto \over \sim \;$}}

\begin{document}

\title{Phase-Space Evolution of Dark Matter Halos}

\author{S. Peirani\inst{1,}\inst{2}  \and J. A. de Freitas Pacheco\inst{3} }

    \institute{
               Institut d'Astrophysique de Paris, 98 bis Bd Arago, 75014
               Paris, France - UMR 7095 CNRS
               - Universit\'e Pierre et Marie Curie\\
               \email{peirani@iap.fr}
        \and
	       Department of Physics, University of Oxford,
               Denys Wilkinson Building,
               Keble Road, Oxford OX1 3RH, UK
        \and
 	       Observatoire de la C\^ote d'Azur -
               Laboratoire Cassiop\'ee - UMR 6202 -
               BP 4229 - 063
               04 - Nice Cedex 4 - France\\
               \email{pacheco@oca.eu}
             }

   \date{Received ..., ...; accepted ..., ...}

 
  \abstract
   {
In a Universe dominated by dark matter, halos are continuously accreting
mass (violently or not) and such mechanism affects their dynamical state.
   }
   {
The evolution of dark matter halos in phase-space,  and using
the phase-space density indicator $Q=\rho/\sigma^3$ as a tracer, is discussed.
   }
   {
We have performed cosmological $N$-body simulations from which we have carried
a detailed study of the evolution of $\sim 35$ dark halos in the interval
$0\leq z \leq 10$.
   }
   {
The follow up of individual halos indicates two distinct evolutionary 
phases. First, an early and fast decrease 
of $Q$ associated to virialization after the gravitational collapse
takes place. The nice 
agreement between simulated data and theoretical expectations based on 
the spherical collapse model support such a conjecture. The late and long period where 
a slow decrease of the phase-space density occurs is related to accretion and
merger episodes. The study of some merger events in 
the phase-space (radial velocity versus radial distance)
reveals the formation of structures quite similar to caustics generated in 
secondary infall models of halo formation. 
After mixing in phase-space, halos 
in quasi-equilibrium have flat-topped velocity 
distributions (negative kurtosis) with respect 
to Gaussians. The effect is more noticiable for 
captured satellites and/or substructures
than for the host halo.
   }
   {}

 \keywords{dark matter -- Galaxies: halos -- Galaxies: interactions --
Methods: N-body Simulations
               }

  \maketitle   
%

\section{Introduction}

In the cold dark matter paradigm, galaxies are formed when baryonic gas 
falls into the gravitational
potential well of cold dark matter (CDM) halos. These halos evolve by  
accreting mass either by  quasi continuous
processes or by merger events, when their masses vary sudden and significantly. Mass 
accretion is also an important mechanism by which halos acquire angular momentum after 
turnaround, since during the collapse phase tidal torques become inefficient  
(Peirani et al. 2004). Moreover, Wechsler et al. (2002)
found an important correlation between accretion and the halo concentration.
Halos in a state of high accretion have central densities related
to the background density while those with low infall rates have approximately
constant central densities.

After a merger episode, the resulting halo is not in equilibrium. Rapid 
variations in the gravitational potential
contribute to the relaxation of the resulting system through a well known 
mechanism dubbed ``violent 
relaxation'' (Lynden-Bell, 1967). This process, after a few dynamical time 
scales ($t_{dyn} \sim 1/\sqrt{G\bar\rho}$),
produces a smooth mass-independent distribution function (DF) as a result of the 
gravitational scattering of particles.
Violent relaxation leads to a more mixed 
system (Tremaine et al. 1986), reducing the value of the
coarse-grained DF. Mixing effects were also considered by Dehnen (2005), who
has investigated the behavior of the {\it excess-mass function $D(f)$} defined
in a phase-space volume where the coarse-grained distribution
function is greater than a given value $f$. For central density profiles represented
by a power law ($\rho\propto r^{-\gamma}$), Dehnen (2005) found that 
$D \propto f^{-2(3-\gamma)/(6-\gamma)}$, suggesting that steeper cusps are less
mixed that shallower ones. As a corollary, if halos having different power law 
density profiles merge, the resulting halo cannot have a cusp steeper than those
of the progenitors. A similar approach was adopted by Arad et al. (2004), who 
defined instead the function $v(f)$ such as $v(f)df$ be the phase-space volume occupied
by phase-space elements where the density lies in the range $f, f+df$. They have
found from cosmological simulations that $v(f)$ is quite well described by a
power law, e.g., $v(f) \propto f^{-2.5\pm 0.05}$, over three to five decades in $f$.
According to them, such a power law behavior reflects the halo substructure, consequence
of the hierarchical clustering and not the result of violent relaxation.   

Investigations on the resulting equilibrium state after a merger episode, based 
on numerical simulations, use
in general a more simple estimator for the coarse-grained DF defined 
as $Q = \rho/\sigma^3$, where $\rho$ and $\sigma$
are respectively the density and the cube of the 1-D velocity dispersion of 
dark matter particles inside a considered
volume. High resolution simulations of galaxy-size CDM halos indicate an 
increase of $Q$ towards the center
(Taylor \& Navarro, 2001) and similar results were obtained 
by Boylan-Kolchin \& Ma (2004), Rasia et al. (2004),  Ascasibar et al. (2004),
Dehnen \& McLaughlin (2005),  Hoffman et
al. (2007), Ascasibar \& Gottl{\"o}ber (2008), Vass et al (2008a, 2008b),
 who have
also obtained a power-law variation, e.g., $Q \propto r^{-\beta}$  for 
cluster-size halos, with $\beta$ quite close
to the value found by Taylor \& Navarro, namely, $\beta \approx$ 1.87. 
Moreover, similar trends have been found by Knollmann, Knebe \& Hoffman (2008) 
considering different cosmogonies although they found that the slope $\beta$ 
depends on the concentration parameter of dark matter halos.
No adequate explanation presently exists
for such a power-law profile which describes  the phase-space density profile in dark
halos. However, Austin et al. (2005)
used semi analytic extended secondary infall models to show that such a behavior 
is a robust feature of
virialized halos, which have attained equilibrium via violent relaxation
and not the result of hierarchical merging, a conclusion in contradiction with
that by Arad et al. (2004).

Analyses of the phase-space density in the core of dwarf spheroidal galaxies, rotating 
dwarfs, low surface brightness
galaxies and clusters suggest another 
scaling law, e.g., $Q \propto \sigma^{-n}$, with $n \sim$ 3-4 (Dalcanton
\& Hogan, 2001). This scaling can be understood if the merging halos were 
initially close to equilibrium and
if the fusion process preserves approximately the physical density as 
each layer is homologously added to
form the new system (Dalcanton \& Hogan, 2001). A scaling relation 
close to $Q \propto \sigma^{-3}$ was
obtained from cosmological simulations by Dav\'e at al. (2001) for collisionless 
as well as
for self-interacting dark matter (SIDM). This result for SIDM is unexpected since 
in this case the material should be
compressed to higher densities during a merger event, sinking to where it 
reaches the local pressure
equilibrium and where the specific entropy matches. Therefore mergers should 
occur at nearly constant $Q$ rather than constant density.

In a previous investigation (Peirani et al. 2006, 
hereafter PDP06), we have 
reported cosmological simulations aiming to study the evolution 
of the phase-space density 
indicator $Q$ in core of dark matter halos. Halos were classed in two 
different catalogs:  the {\it accretion} , comprising 
781 objects which have never undergone a major merger event and whose 
masses varied continously and smoothly. The 
{\it merger catalog} contains 567 halos which had at least one major 
merger event, corresponding to an 
increase of their masses at least by a factor 1/3. These simulations 
indicate that the phase-space density 
decreases continuously in time, scaling with the velocity dispersion 
as $Q \propto \sigma^{-2.1}$ and with the 
halo mass as $Q \propto M^{-0.82}$. No differences in these scaling relations 
were seen between  ``cold" and ``warm" 
dark matter models but halos which have underwent 
important merger events are, on the average, more relaxed  
having $Q$ values lower than halos of 
the accretion class. The follow up of individual halos indicates an early and
fast phase in which $Q$ 
decreases on the average by a factor of 40 followed by a long period in 
which $Q$ further decreases by about factor 
of 20. The decrease of $Q$ (or the {\it increase} in the entropy) in the 
first phase is probably a consequence 
of the randomization of bulk motions during the first shell crossing while 
accretion and merger events are responsible 
for the slow decrease observed in late epochs.

In the present work we report a detailed investigation of 35 CDM halos whose 
evolution was followed from $z \sim 10$
up today. For each of these halos, we have estimated the redshift at which 
the first shell crossing occurs and which
we assume to coincide with ``virialization". The phase-space density 
indicator $Q$ in core of these halos was estimated 
at that moment and compared with theoretical estimates performed through 
the spherical model. The agreement between
theoretical and numerical estimates confirms our previous conjecture 
concerning the rapid decrease of $Q$ observed
in the early evolutionary phases. We have also studied the late behavior 
in some specific examples to show 
the evolution of the velocity distribution during a merger event and 
the evolution of structures in the phase-space.
This paper is organized as follows: in Section 2 we describe 
briefly the simulations, in Section 3 the phase-space
density at ``virialization" is discussed, in Section 4 we describe 
some merger episodes and the behavior of the
velocity distribution as well as of structures in phase-space 
and, finally, in Section 5 we present our
main conclusions.

\section{The simulations}

In the present work we have used the same halo catalogs as in PDP06, including 
objects with masses in the
range $10^{10}-10^{13}\,M_{\odot}$. For the sake of completeness, we summarize 
here the main steps
performed to prepare these catalogs.

The N-body simulation uses the adaptive particle-particle/particle-mesh ($AP^3M$) 
code HYDRA (Couchman et al. 1995). The adopted cosmological parameters 
were h = 0.65, $\Omega_m = 0.3$ and
$\Omega_{\Lambda} = 0.7$, with the power spectrum 
normalization $\sigma_8 = 0.9$. The simulation was
performed in a box of side 30$h^{-1}$ Mpc, including 256$^3$ 
particles, corresponding to a mass resolution
of $2.05\times 10^8\,M_{\odot}$. The simulation started at $z=50$ and 
ended at the present time ($z=0$).
Halos were initially detected by using a friends-of-friends (FOF) 
algorithm (Davis et al. 1985) and, in a second step, unbound
particles were removed by an iterative procedure. Thus, our selected halos 
are all gravitationally bound
objects. In the total, 1348 halos were detected, constituting the two aforementioned 
catalogs (PDP06).
The evolution of these objects was chased from $z=3.5$ until $z=0$, but 
about 40 halos with initially enough
particles (N$\geq$50) were followed during a longer time 
interval ($10\leq z \leq 0$). For each halo, we have monitored the evolution of the 
mass, density, virial ratio and velocity dispersion.

\section{The phase-space density at virialization}

According to the analysis by PDP06, the phase-space density $Q$ decreases 
continuously as halos accrete matter. The derivative $dQ/dt$ shows the 
presence of ``valleys" which coincide with peaks
in the derivative $d\sigma/dt$ of the velocity dispersion. These structures are 
always associated
with sudden variations in the mass, induced by merger events and leading 
to a slight
``heating", followed by a relaxation of the system, measured by the decrease in 
the phase-space density indicator $Q$, which is equivalent to say that the 
entropy of the system
increased. If this behavior characterizes in general the late evolution of 
halos, the first and
strong peak in the derivative of the velocity dispersion, correlated with the 
first and deep valley in
the derivative of the phase-space density (see Fig.~4 in PDP06), is not a 
consequence of merging, but probably a consequence of the gravitational 
collapse. If this interpretation is correct, an important
energy transfer from bulk to random motions should occur due to collective 
effects, heating all the particles, regardless their initial energies.

In order to test our conjecture, we have first derived for a sample of 35 halos, 
which have been followed in the interval $10\leq z \leq 0$, which includes the redshift 
at which the first shell crossing
occurs. We assumed here, as in Sugerman et al. (2000), that 
the first shell crossing episode coincides
with the time of maximum velocity dispersion as well as with ``virialization''. In 
fact, as halos
accrete mass, the virial ratio $2T/\mid W\mid$ approaches asymptotically the 
unity and, just
after the first shell crossing, they are only in quasi dynamical equilibrium. For 
our subsequent
analysis, we assume for simplicity that the relaxation process was almost achieved.

\begin{figure}
\rotatebox{0}{\includegraphics[width=\columnwidth]{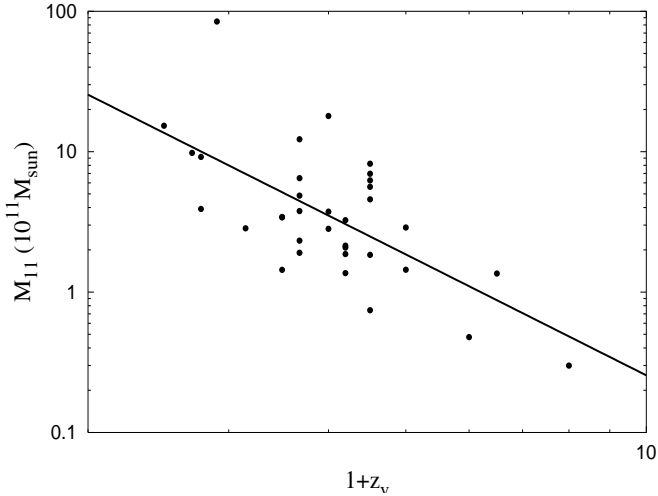}}
\caption{The halo mass as a function of the redshift $z_v$ at which ``virialization'' 
occurs (defined as the time of maximum velocity dispersion). The solid line
represents the best fit solution.}
\end{figure}

\begin{figure}
\rotatebox{0}{\includegraphics[width=\columnwidth]{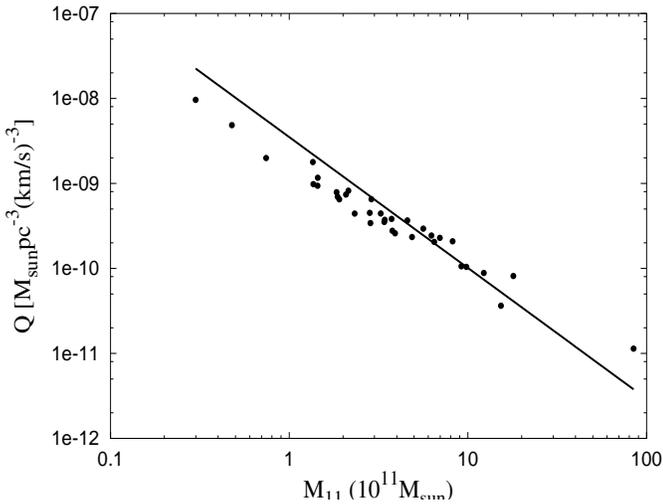}}
\caption{The phase-space density Q as a function of the halo mass. The points
represent values derived from simulated data whereas the solid line is obtained
from relation (8).}
\end{figure}

In fig. 1, the halo mass is plotted as a function of the redshift $z_v$ at
which ``virialization'' occurs.
Notice that, as expected in the hierarchical picture, less massive halos collapse 
first and that
large galaxy-size halos ($\sim 10^{12}\,M_{\odot}$) are being
virialized around $z_v \sim 1-2$.
A best fit of these simulated data gives
\begin{equation}
M_{11}=\frac{185}{(1+z_v)^{2.86}}
\end{equation}
where $M_{11}$ stands for the halo mass in units of $10^{11}\,M_{\odot}$.

The phase-space density $Q$ at ``virialization'' was calculated  
as follows: for each halo, the gravitational
radius $r_g = GM/\mid W\mid$ was computed as well as the mean 
density $\bar\rho = 3M/(4\pi r_g^3)$. The
1-D velocity dispersion $\sigma$ was calculated by assuming isotropy, e.g.,
$\sigma = \sqrt{(\sigma_x^2 + \sigma^2_y + \sigma^2_z)/3}$ and,
 finally the phase-space density indicator, $Q=\bar\rho/\sigma^3$.

In order to perform a theoretical estimate of the value of the phase-space 
density $Q$ just after
virialization, we have appealed to the spherical collapse model, including 
effects of a cosmological constant, in order to be consistent with the cosmology
adopted in our simulations.

The expected density contrast at virialization, using the results of
Bryan \& Norman (1998), is
approximately given by
\begin{equation}
\Delta_v = 18\pi^2+82\lbrack \Omega_m(z_v)-1\rbrack - 39\lbrack \Omega_m(z_v)-1\rbrack^2
\end{equation}
Thus, the mean halo density at virialization is
\begin{equation}
\rho_v =\Delta_v\frac{3H_0^2}{8\pi G}\Omega_m^0(1+z_v)^3=\frac{3M}{4\pi R_v^3}
\end{equation}
where $\Omega_m^0$ is the present matter (dark+baryonic) density parameter and $R_v$ is
the radius at virial equilibrium. 

Defining the ratio between the virialization and the turnaround 
radii as $\eta=R_v/R_0$ and using the 
energy conservation as well as the virial relation, one obtains (Lahav et al. 1991)
\begin{equation}
\lambda\eta^3-2(2+\lambda)\eta+2=0
\end{equation}
where we have introduced the parameter $\lambda = \Omega_vH^2_0R^3_0/GM$.
The solution of this cubic equation can approximately be expressed by
\begin{equation}
\eta = 0.5 - 0.20936\lambda + 0.04949\lambda^2
\end{equation}
Since the ratio between densities at turnaround and 
virialization is $\rho_{ta}/\rho_v = \eta^3$,
using eq.(3) and the definition of the parameter $\lambda$ one obtains
\begin{equation}
\lambda\eta^3= \frac{3\Omega_vH^2_0}{4\pi G\rho_v}
\end{equation}

On the other hand, since $3\sigma^2=-2E$, where $E$ is the total energy, we 
can express the
1-D velocity dispersion as
\begin{equation}
\sigma^3=\lbrack\frac{2}{3\lambda^{1/3}}+
\frac{\lambda^{2/3}}{3}\rbrack^{3/2}(\Omega_vH_0^2)^{1/2}GM
\end{equation}

Therefore, the procedure adopted to estimate $Q$ was: for a given virialization 
redshift $z_v$, one computes
the virialization density $\rho_v$ from eqs.~(2) and (3). Then, using eqs.~(5) and (6) 
the value of $\lambda$
at the considered $z_v$ is derived. This value is used in eq.~(7), together with
the mass $M$ virializing
at $z_v$ and estimated from eq.~(1), to obtain the velocity 
dispersion. The resulting
values of the phase-space density $Q$ are quite well fitted by the relation
\begin{equation}
Q \approx \frac{3.51\times 10^{-9}}{M_{11}^{1.54}} \,\, M_{\odot}pc^{-3}km^{-3}s^{-3}
\end{equation}

Figure 2 compares the expected values of the phase-space density $Q$ 
as a function of the halo mass, computed from the equation above and
values derived from simulated data. The agreement is quite good in spite of the 
fact that the model
assumes a constant halo mass, but confirms our previous conjecture that the 
early and fast decrease
of the phase-space density observed in the halo history reflects 
the ``thermalization'' of bulk motions.

\section{Phase-Space Mixing in Mergers}

In the late evolutionary phases of dark matter halos, the rate at which the 
phase-space density
decreases depends on the frequency of mergers and on the amount of mass accreted in
these events. Subhalos captured in quite eccentric orbits are generally 
disrupted completely by tidal forces, transferring angular momentum to 
the host halo and thus increasing its spin (Peirani et al. 2004) but forming, 
as we shall see latter, streams detectable
in phase-space diagrams before to mix completely with the background material.

An attempt to verify whether the merging history of our Galaxy left ``finger-prints'' in the
phase-space structure of nearby stars was made by Helmi \& White (1999). In
that investigation,
the infall of satellites onto a fixed potential was followed numerically as well as the
evolution of the debris in phase-space. Helmi \& White found that 
after $\sim$ 10 Gyr, stars
having a common origin are distributed smoothly in space, but form clumps 
in velocity space.

We have searched in our simulation for merger episodes involving massive 
halos, e.g., having
a mass greater than $10^{12}\,M_{\odot}$ at $z=0$, in order to perform evolutionary studies
in the phase-space.

The first example consists of a main halo having presently a mass of
about $4.0\times 10^{12}\,M_{\odot}$,
which was caught in the act of capture of two subhalos (at $z = 0.92$) with masses at
that moment respectively
equal to $1.6\times 10^{11}\,M_{\odot}$ (subhalo-1) 
and $2.8\times 10^{11}\,M_{\odot}$ (subhalo-2).
The orbits of each subhalo are sketched in fig. 3. The trajectory of the 
center of mass of subhalo-1
is represented both in the orbital and in the transversal planes in the first 
two panels. The
trajectory of subhalo-2 is represented in the orbital plane up to $z = 0.48$ 
only, since its identity
is completely lost at late times. Subhalo-1 has an impact parameter or
orbital angular momentum
higher than the subhalo-2, illustrating the well known effect of this dynamical parameter
on the tidal stripping of captured objects.

\begin{figure*}
\rotatebox{0}{\includegraphics[width=18.0cm]{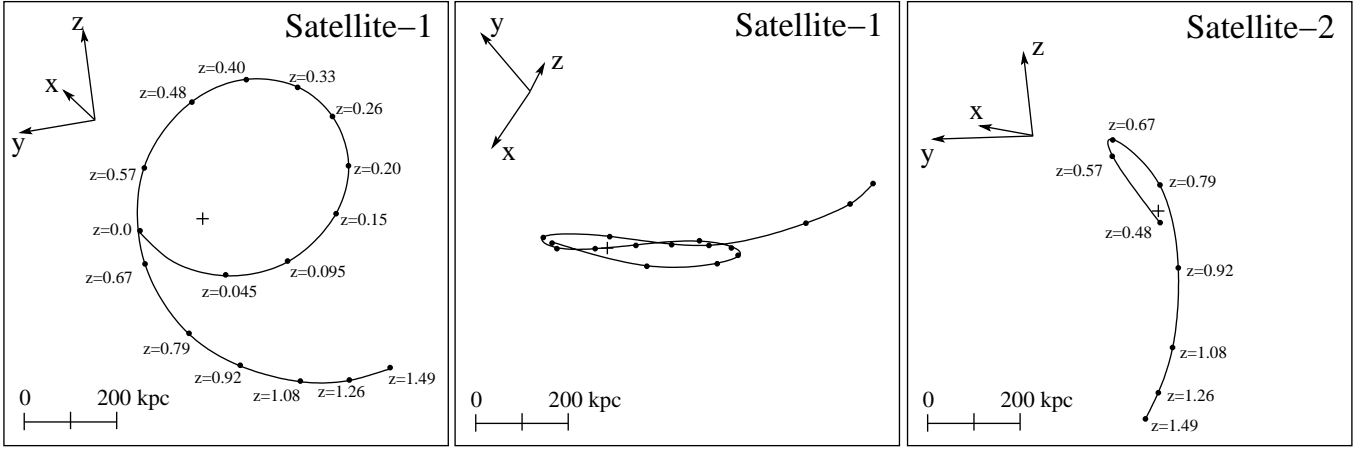}}
\caption{Sketch of the orbits of satellites in 3D-space for example 1. In the left 
and right panels, the motion in the orbital plane is represented for both satellites,
 while in the middle panel the motion of satellite-1 is represented in the
transversal plane. The cross indicates the position of the center of mass
 of the system.}
\end{figure*}

Successive snapshots of the phase-space evolution (radial velocity versus 
radial distance) are
shown in fig.~4. The subhalo-1 has an orbital period of about 6 Gyr and its 
periastron is about
100 kpc. Tidal forces strip off only the outer and loosely bound particles, the 
object preserving
most of its identity after 8 Gyr. However, already at $z \approx$ 0.67 
(near periastron), stripped
particles with high positive velocities can be seen, while the bulk has 
negative values. Effects
in the velocity distribution are better visualized in fig.~5, where the 
corresponding radial
velocity distributions are shown for each snapshot of fig.~4. Initially ($z = 0.92$) 
three distinct velocity distributions can be identified, associated respectively to the 
main halo and the two subhalos, with velocity
dispersions of $\sigma_H$ = 202 km/s, $\sigma_1$ = 51 km/s
and $\sigma_2$ = 78 km/s. For comparison, the best fitted Gaussians are also shown
as solid lines.
In spite of subhalo-1 to have preserved its identity, stripped 
particles mix in the velocity space and
this process can be followed in the different snapshots 
either in fig.~4 or fig.~5. Notice in particular
the appearance of extended tails in the velocity distribution of the 
captured halos. Subhalo-2
has a short orbital period ($\sim$ 2 Gyr) and a quite eccentric orbit. 
Extended tidal arms are developed after the various periastron passage,
forming structures spatially displaced by the orbital decay resulting from 
dynamical friction. In this case, the mixing
in the velocity space (see, fig.~5) is quite efficient and after 3.4 Gyr 
the particle velocity distribution of subhalo-2 coincides practically with 
that of the main halo.

\begin{figure*}
\rotatebox{0}{\includegraphics[width=18cm]{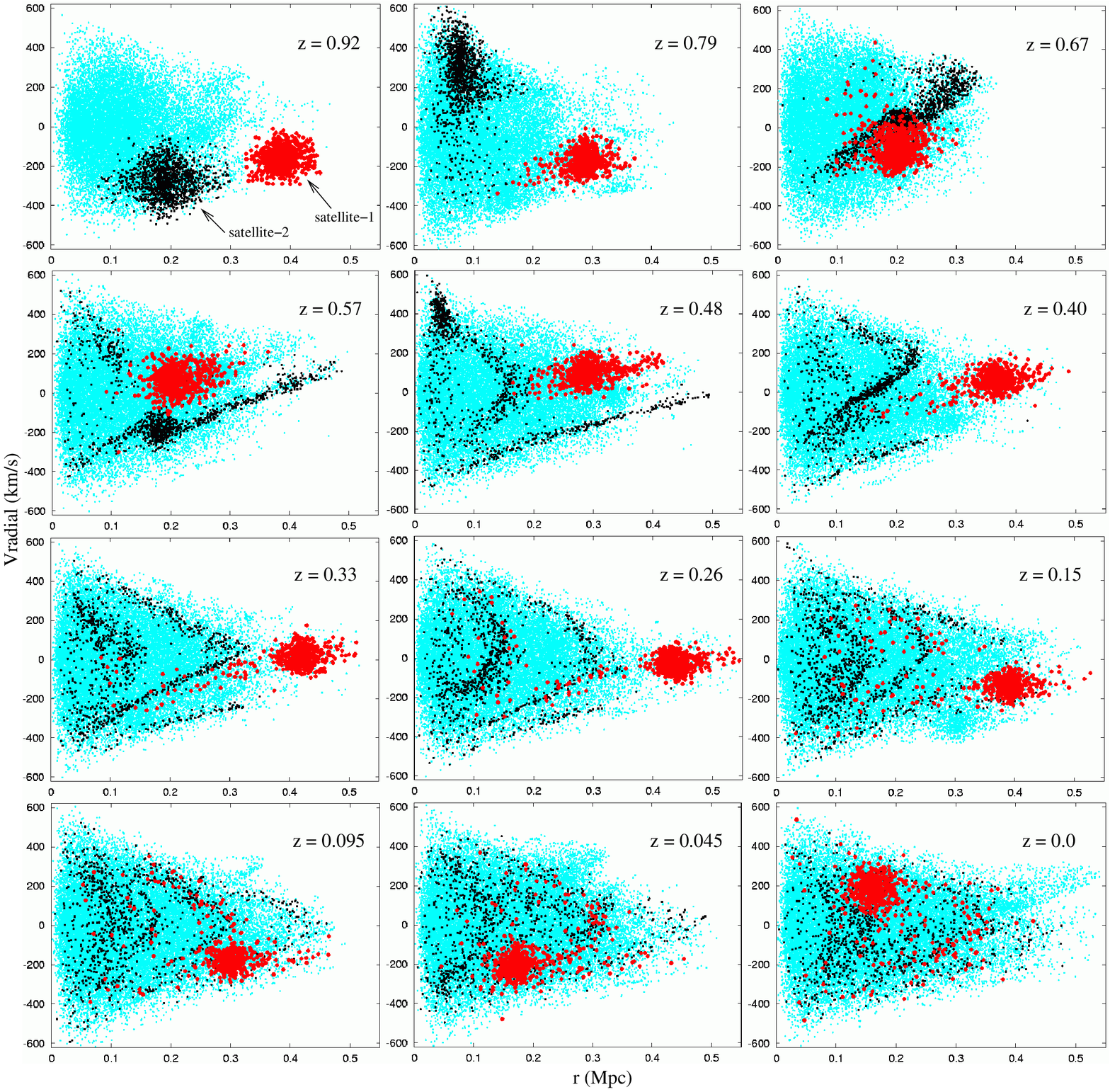}}
\caption{The phase-space evolution (radial velocity versus radial distance)
for the host halo and satellites of example 1. Upper right labels
indicate the redshift.
Satellite-1 and Satellite-2 are respectively represented by red and black points.}
\end{figure*}

In spite of Gaussians be able to fit reasonable well the radial velocity
distribution of the main halo (mean determination coefficient $r = 0.97$)
and satellites (but $r = 0.86$ for subhalo-2),
we have also calculated the kurtosis, $k = (<v^4>/<v^2>^2)-3$, for both the
main halo and the subhalo-2 at $z=0$. For the main halo, the kurtosis was
evaluated for ten shells with 2800 particles each, while for subhalo-2 we have
considered only three shells with about 400 particles each. No particular trend
with the radial distance was found, with the main halo having a mean 
kurtosis $k = -0.57$ and the subhalo-2 $k = -0.73$. These values indicate
that the radial velocity distribution is flat-topped with respect to a Gaussian
and that the effect in more accentuated in the resulting substructure. 

Another interesting case of capture is illustrated by our second example. Here the main halo
has a mass of about $3.3\times10^{12}$ M$_{\odot}$ and the capture of a satellite with a
mass of $5.2\times10^{11}$ M$_{\odot}$ occurs at $z = 0.92$. Here again masses correspond
to that instant, since at the end, due to the continuous infall process, masses are
substantially higher. Fig.~6 shows snapshots of the phase space
at different redshifts. Structures are gradually formed as the satellite passes by periastron. The
observed streams in the phase-space are very similar to those derived from models of halo
formation by secondary infall (see, for instance, Sikivie et al. 1997 and references therein).
However, structures seen in the latter case are formed by successive passages of infalling 
(or outgoing) particles
and are associated to ``caustics''. From a mathematical point of view, caustics have an infinite
density. In physical systems, the density is limited by the fact that the motion of DM particles
is not purely radial (non-zero angular momentum) and by a non-zero velocity dispersion
(Sikivie et al. 1997). If the nature of DM particles is specified, then other constraints
can be imposed. For instance, neutralinos and anti-neutralinos may annihilate with a rate proportional
to the square of their density, a mechanism which imposes upper bounds to the density. Moreover,
neutralinos are fermions and one would expect that a ``classical'' behavior must be abandoned
when their de Broglie wavelength will be greater than the mean inter-particle separation. This occurs
for densities {\bf n} satisfying $n > (m_{\chi}v/h)^3$, when repulsive forces of quantum nature
due to the Pauli principle become effective, introducing a further physical mechanism able
to impose limits on the density. In a given instant and for a fixed value of the 
radial distance, a certain number of velocity peaks, corresponding
to different streams can be seen in the phase-space diagram (Fig.~6), which depends on the
initial angular momentum, as simulations suggest. In the extreme limit of circular orbits only one
stream should be observed, a case comparable to that of satellite-1 
in the first example. The width $\delta v_m$ of the
m-velocity peak associated to the m-caustic is generally estimated 
from the Liouville theorem,
using the velocity dispersion and density at the moment of their first turnaround.
In general and in the present case, this procedure can not be 
applied. The fine-grained DF is
strictly conserved in the phase-space flow but this is not the case of the coarse-grained DF,
since phase-space elements of high density are stretched out and folded
with elements of low density because of mixing (Lyndell-Bell 1967).

\begin{figure*}
\rotatebox{0}{\includegraphics[width=18cm]{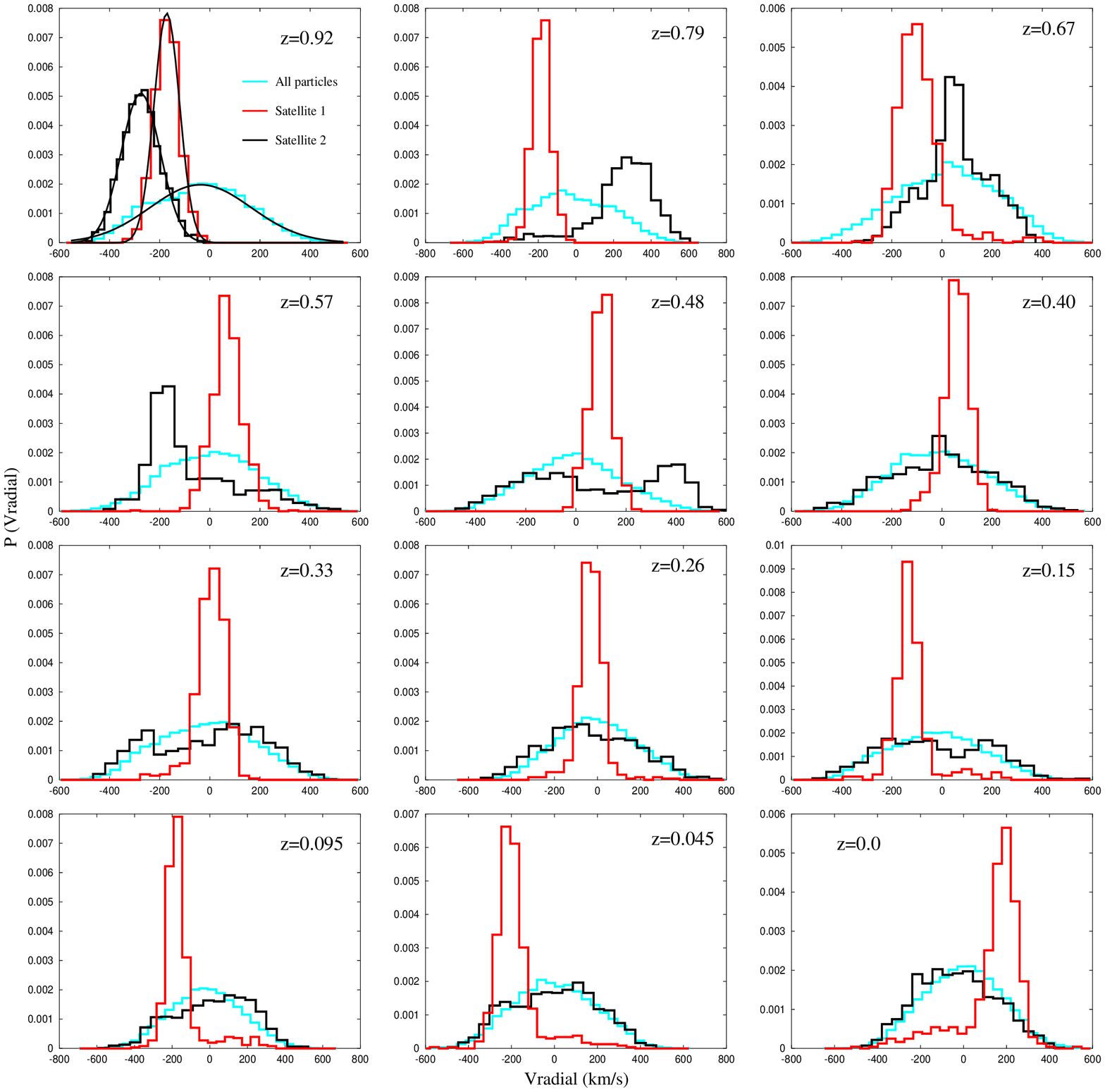}}
\caption{The evolution of the radial velocity distribution for the host halo
(light blue), satellite-1 (red) and satellite-2 (black) for example 1. At 
$z=0.92$, the best fitted Gaussian distribution is superimposed to histograms.
Notice that the velocity distribution of satellite-1 is preserved but it develops
 high velocity tails.}
\end{figure*}

\begin{figure*}
\rotatebox{0}{\includegraphics[width=18cm]{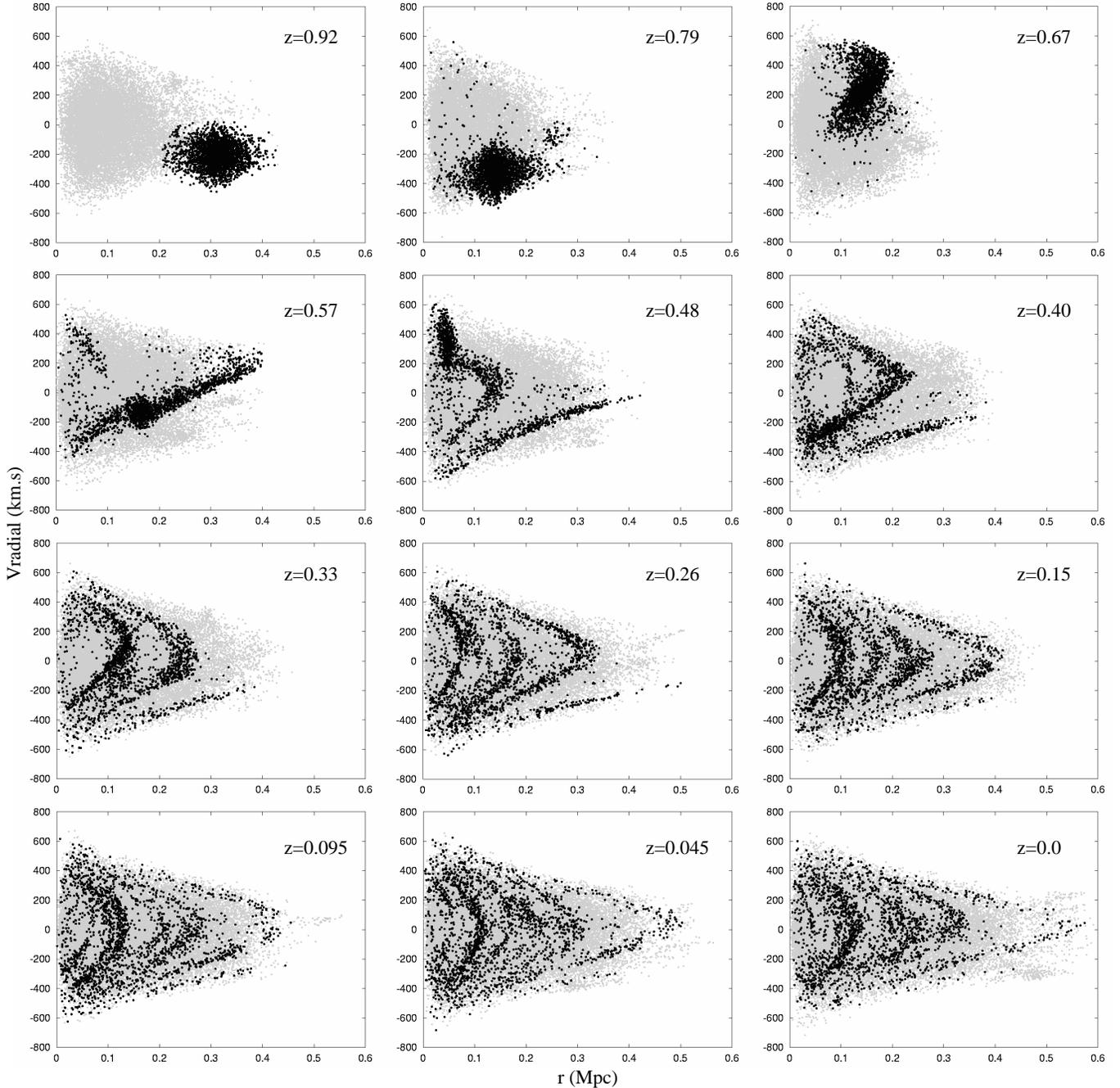}}
\caption{The phase-space evolution of the host halo and  satellite (black points)
for example 2. The latter is  completely disrupted by tidal forces. The different 
snapshots show the formation of streams in phase-space.}
\end{figure*}

\begin{figure}
\rotatebox{0}{\includegraphics[width=\columnwidth]{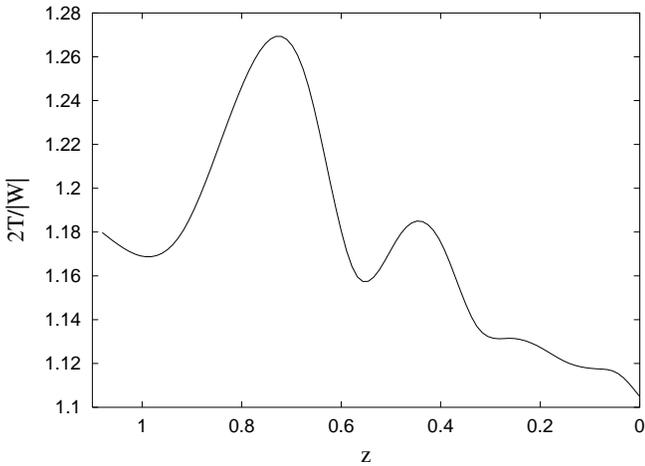}}
\caption{The evolution of the virial ratio $2T/|W|$ as a a function of the redshift (example 2).
 Each maximum corresponds to the passage of the subhalo by periastron.}
\end{figure}

\begin{figure}
\rotatebox{0}{\includegraphics[width=\columnwidth]{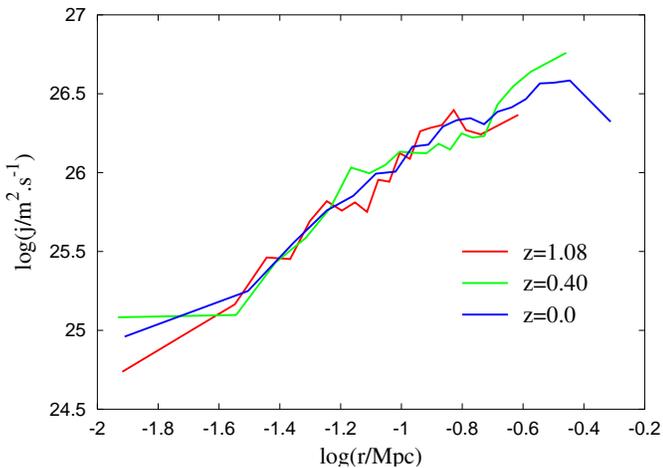}}
\caption{The radial profile of the specific angular momentum $j$ for the
merger example 2. Consequence of a "head-on" collision, no significant
modifications in the profile of $j$ is observed after the merger event
($z=0$) which occurred at $z=1.08$.
}
\end{figure}

\begin{figure*}
\rotatebox{0}{\includegraphics[width=18cm]{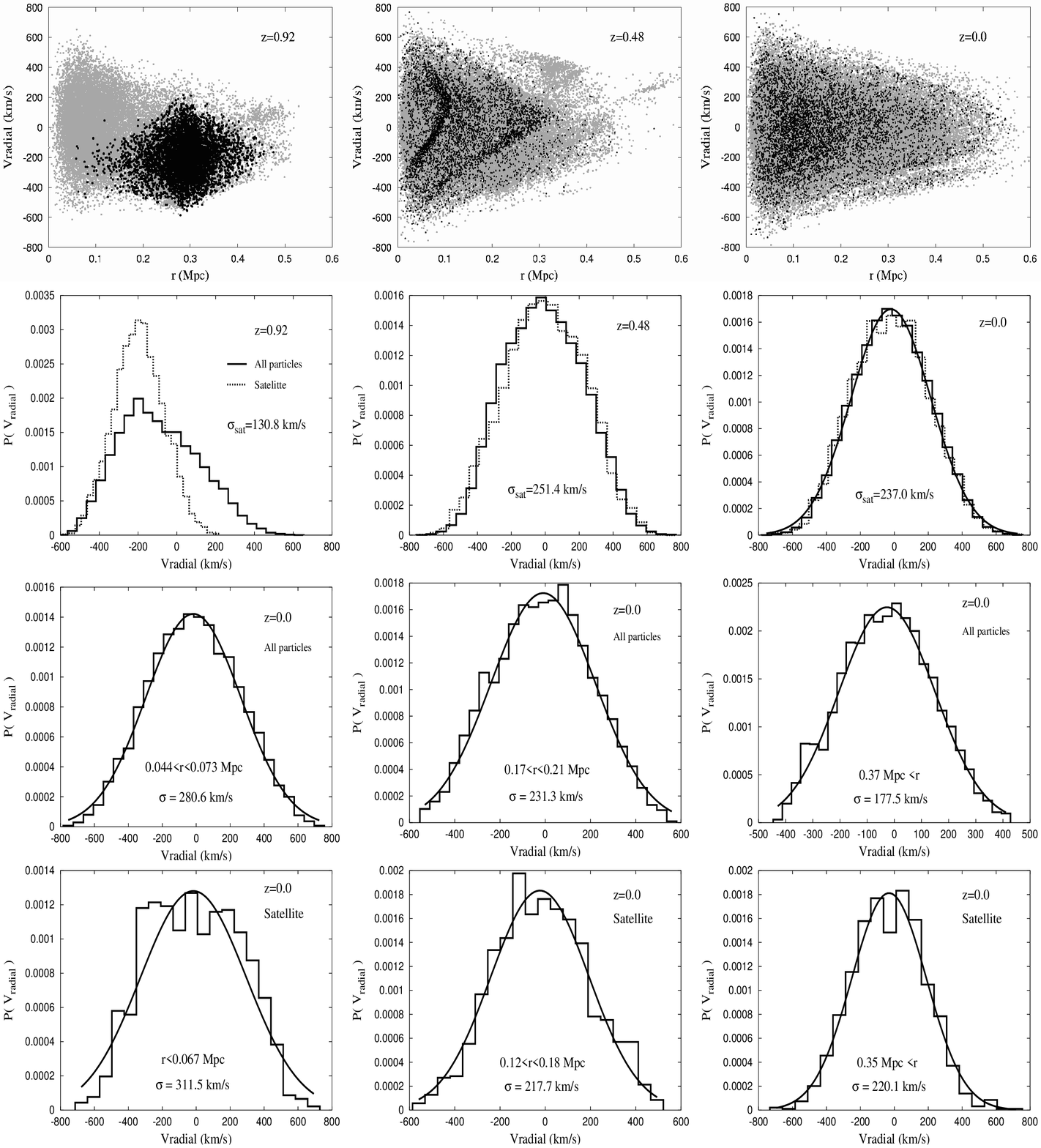}}
\caption{(Top) The phase-space evolution of the main halo and a satellite (black points)
completely disrupted by tidal forces, in example 3.
The other panels show respectively the evolution of
the radial velocity distribution for all particles of both host and satellite (second line),
and the radial velocity distribution at different shells from the center (all particles, third line
and for the satellite, fourth line). Solid lines represent the best Gaussian fit.}
\end{figure*}

It's worth mentioning that the time evolution of the virial ratio $2T/|W|$ 
of the whole system 
has clear maxima corresponding to successive pericenter passages of the center of mass of 
the sub-halo (see Fig.~7), in agreement with merger simulation 
by Valluri et al. (2007). However, in 
this particular case, there is no expressive redistribution of the specific angular 
momentum {\bf j} after the merger event, the
radial profiles of {\bf j} being practically the same before 
and after merging (see fig.~8), since 
the collision orbit is almost radial and, consequently, no 
significant transfer of angular momentum occurs. 
As in the previous example, the analysis of the kurtosis at $z = 0$ indicates 
top-flatted velocity
distributions with the satellite-1 having a flatter 
distribution (mean kurtosis $k = -0.85$ in
comparison with $k = -0.64$ for the main halo).

In the third example of a merger episode, the main halo has a
mass of $4.3\times 10^{12}\,M_{\odot}$ and underwent
an important fusion at z=0.92, since the captured object has a 
comparable mass ($\sim 1.2\times 10^{12}\,M_{\odot}$).
In the three upper panels of fig.~9, the phase-space evolution is
shown at three distinct redshifts. At $z=0.48$, structures formed
at different periastron passages can be seen, while at $z=0$ the 
captured halo is completely disrupted and its particles are well mixed
in phase-space. This behavior is consistent with the simulations by
Valluri et al. (2007), which indicate that mixing in the space of the dynamical 
variables $(E,J)$ occurs primarily during periastron passages and are
driven probably by compressive tidal shocks.
The subsequent panels show the evolution of the radial
velocity distribution. The second line shows the evolution for all
particles of the main halo and satellite, which at $z=0.92$ has a velocity dispersion
of 130 km/s. At $z=0.48$ and $z=0$ both velocity distributions practically match.
The average kurtosis are $k= -0.57$ and $k = -0.65$ for the main halo and satellite
respectively, confirming the trend that substructures have top-flatted velocity
distributions more accentuated than those of the main halos.

Notice that the fact that all selected merger examples
occur at $z = 0.92$ is completely fortuitous and is probably associated to
a maximum rate of mergers around $z \sim 1$ as well as to the time resolution
adopted in the present study. 

The main halos in the aforementioned examples have typically $(3-4)\times 10^4$
particles while for satellites the numbers are about one order of magnitude lower. In the simulations by Wojtak et al. (2005), cluster size halos with few times $10^4$ particles (numbers comparable to those of our selected examples)
were considered. These authors concluded that the velocity distribution of
main halos are only Gaussian to a good approximation near the center, but more and more flat-topped with respect to Gaussians when approaching the 
virial radius.
Numerical experiments of higher resolution were performed by Diemand et al. (2004) and Kazantzidis et al. (2004). The former authors considered galactic
and cluster size halos extracted from cosmological simulations, including
several millions of particles. They concluded that subhalos have flat-topped
velocity distributions with a typical kurtosis $k \simeq -0.7$. Equilibrium
halos with about $10^5$ particles were generated by Kazantzidis et al. (2004)
using the Eddington method. They found also that the velocity distribution of such systems are less peaked than Gaussians.

In spite of the different employed resolutions, all these 
investigations (includind our own) lead to consistent results, e.g., the
velocity distribution of halos are flat-topped with kurtosis in the
range $-0.4 > k > -0.9$.

\section{Conclusions}

The dynamical evolution of dark matter halos after the gravitational 
collapse depends on their accretion history. The phase-space density 
indicator $Q$ decreases by a large factor ($\sim 40$) 
during the first shell crossing
as a consequence of the randomization process of initial bulk motions. Our 
theoretical estimates of the
phase-space density $Q$ based on the spherical collapse model are in agreement 
with our simulated data, suggesting
that at ``virialization" the phase-space density scales with the halo 
mass as $Q \propto M^{-1.5}$. The analysis
by Armarzguioui \& Gron (2005) indicates that the entropy of an ideal gas 
increases to the first order in
the gravitational collapse within the context of the FRW cosmology. This entropy 
increase is a consequence of
the transfer of gravitational potential energy to thermal energy during the 
collapse, a process similar
to that observed for colisionless matter. Effects of shell crossing 
play probably a non negligible role in the way halos reach equilibrium 
and should be taken into account in future studies, since they
affect the resulting  ``virial" radius of the system (S\'anchez-Conde et al. 2006).   

Non-disrupted satellites develop high velocity tails in their radial 
velocity distribution. The mecanism (or mechanisms) responsible 
for such a heating is not well 
established. Funato, Makino \& Ebisuzaki (1992)
based on N-body simulations of collisionless systems argued 
that wave-particle interactions, e.g., collective
effects driven by large fluctuations of the gravitational potential are 
able to produce a substantial
heating in a few dynamical time scales, regardless the particle initial 
energy. Another possibility
is the heating produced by compressive tides that arise when one 
collisionless gravitating system passes
through another on a time scale shorter than the internal dynamical time 
of the infalling object. Such
tides can impulsively heat particles as a result of the transient deepening 
of the net gravitational
potential (Valluri et al. 2007). Impulsive compressive tidal shocks are 
well known to produce changes in the internal structure of globular 
clusters (Gnedin et al. 1999) or in subhalos orbiting within
massive hosts (Kravtsov et al. 2004). 

Disrupted satellites leave ``finger-prints'' in phase-space, generating streams 
which depend on the initial
orbital angular momentum and which remind caustic structures seen in secondary 
infall models of halo
formation.
Future high resolution simulations will certainly give a better
insight on the evolution of these structures.
 It is worth mentioning that extended secondary infall models are also
able to explain the scaling-free 
$Q$ profile (Austin et al. 2005). Stars behave as a colisionless fluid similar to
dark matter. In this
case, one should expect that disrupted satellites will form not only dark matter 
streams but also
stellar streams, which could eventually be detected in phase-space by the 
forthcoming space mission GAIA as also suggested by Brown et al. (2005).

Our previous study (PDP06) and the present work support the view that 
violent variations in the gravitational
potential, which occur in important merger events, lead to more mixed 
systems as the temporal behaviour of the
phase-space density suggests. The study of some merger episodes indicates that
the radial velocity of
completely disrupted satellites and that of the main halo fuse in a common 
distribution in a few dynamical time
scales. Quasi-relaxed halos have top-flatted velocity distributions, which
are observed not only for the bulk of particles but also for particles inside shells
at different distances from the center. The significance of these top-flatted profiles is
measured by the kurtosis (always negative) and this effect is more accentuated
for captured subhalos and/or substructures in comparison with the velocity distribution
of the main halo. Kazantzidis et al. (2004) have developed an algorithm for 
constructing N-body realizations of equilibrium spherical systems, which differs 
from the usual assumption that the local
velocity distribution is Maxwellian. Equilibrium halo models built through their procedure
when evolved, develop velocity distributions less peaked than Gaussians and the disruption
time scale of captured satellites becomes significantly longer in comparison with initial
Maxwellian equilibrium models. Cosmological simulations by Diemand et al. (2004)
lead also to the conclusion that substructures of either galactic-size or cluster-size
halos have top-flatted velocity profiles, in agreement with our results.
 Moreover, an
investigation of structures resulting from head-on collisions of halos having a 
Navarro-Frenk-White density profile was performed by Hansen et al. (2006). They concluded
that the phase-space density indicator $Q$ has a power-law distribution and that
the resulting velocity distribution is not Gaussian, but can be represented by a function
derived from the Tsallis statistics (Tsallis 1988). Thus, deviations from Gaussianity
seems to be a quite general result and, in this sense, the present simulations
represent a contribution to establish an empirical basis for future theoretical developements
in this area.

\acknowledgements
S.\,P.\, acknowledges the financial support from a ANR grant.
S.\,P.\, is grateful to R. Mohayaee for discussions on the early phase of this work.
We also thank  S. Hansen and F. Durier for their
useful comments which have significantly improved this paper.

\end{document}